\documentclass[12pt]{article}
\usepackage{amssymb}
\makeatletter
 
\def\diagram{\m@th\leftwidth=\z@ \rightwidth=\z@ \topheight=\z@
\botheight=\z@ \setbox\@picbox\hbox\bgroup}
 
\def\enddiagram{\egroup\wd\@picbox\rightwidth\unitlength
\ht\@picbox\topheight\unitlength \dp\@picbox\botheight\unitlength
\hskip\leftwidth\unitlength\box\@picbox}
 
\def\bfig{\begin{diagram}}
\def\efig{\end{diagram}}
\newcount\wideness \newcount\leftwidth \newcount\rightwidth
\newcount\highness \newcount\topheight \newcount\botheight
 
\def\ratchet#1#2{\ifnum#1<#2 \global #1=#2 \fi}
 
\def\putbox(#1,#2)#3{%
\horsize{\wideness}{#3} \divide\wideness by 2
{\advance\wideness by #1 \ratchet{\rightwidth}{\wideness}}
{\advance\wideness by -#1 \ratchet{\leftwidth}{\wideness}}
\vertsize{\highness}{#3} \divide\highness by 2
{\advance\highness by #2 \ratchet{\topheight}{\highness}}
{\advance\highness by -#2 \ratchet{\botheight}{\highness}}
\put(#1,#2){\makebox(0,0){$#3$}}}
 
\def\putlbox(#1,#2)#3{%
\horsize{\wideness}{#3}
{\advance\wideness by #1 \ratchet{\rightwidth}{\wideness}}
{\ratchet{\leftwidth}{-#1}}
\vertsize{\highness}{#3} \divide\highness by 2
{\advance\highness by #2 \ratchet{\topheight}{\highness}}
{\advance\highness by -#2 \ratchet{\botheight}{\highness}}
\put(#1,#2){\makebox(0,0)[l]{$#3$}}}
 
\def\putrbox(#1,#2)#3{%
\horsize{\wideness}{#3}
{\ratchet{\rightwidth}{#1}}
{\advance\wideness by -#1 \ratchet{\leftwidth}{\wideness}}
\vertsize{\highness}{#3} \divide\highness by 2
{\advance\highness by #2 \ratchet{\topheight}{\highness}}
{\advance\highness by -#2 \ratchet{\botheight}{\highness}}
\put(#1,#2){\makebox(0,0)[r]{$#3$}}}

\def\adjust[#1]{} 
 
\newcount \coefa
\newcount \coefb
\newcount \coefc
\newcount\tempcounta
\newcount\tempcountb
\newcount\tempcountc
\newcount\tempcountd
\newcount\xext
\newcount\yext
\newcount\xoff
\newcount\yoff
\newcount\gap%
\newcount\arrowtypea
\newcount\arrowtypeb
\newcount\arrowtypec
\newcount\arrowtyped
\newcount\arrowtypee
\newcount\height
\newcount\width
\newcount\xpos
\newcount\ypos
\newcount\run
\newcount\rise
\newcount\arrowlength
\newcount\halflength
\newcount\arrowtype
\newdimen\tempdimen
\newdimen\xlen
\newdimen\ylen
\newsavebox{\tempboxa}%
\newsavebox{\tempboxb}%
\newsavebox{\tempboxc}%
 
\newdimen\w@dth
 
\def\setw@dth#1#2{\setbox\z@\hbox{\m@th$#1$}\w@dth=\wd\z@
\setbox\@ne\hbox{\m@th$#2$}\ifnum\w@dth<\wd\@ne \w@dth=\wd\@ne \fi
\advance\w@dth by 1.2em}
 
\def\t@^#1_#2{\allowbreak\def\n@one{#1}\def\n@two{#2}\mathrel
{\setw@dth{#1}{#2}
\mathop{\hbox to \w@dth{\rightarrowfill}}\limits
\ifx\n@one\empty\else ^{\box\z@}\fi
\ifx\n@two\empty\else _{\box\@ne}\fi}}
\def\t@@^#1{\@ifnextchar_{\t@^{#1}}{\t@^{#1}_{}}}
\def\to{\@ifnextchar^{\t@@}{\t@@^{}}}
 
\def\t@left^#1_#2{\def\n@one{#1}\def\n@two{#2}\mathrel{\setw@dth{#1}{#2}
\mathop{\hbox to \w@dth{\leftarrowfill}}\limits
\ifx\n@one\empty\else ^{\box\z@}\fi
\ifx\n@two\empty\else _{\box\@ne}\fi}}
\def\t@@left^#1{\@ifnextchar_{\t@left^{#1}}{\t@left^{#1}_{}}}
\def\toleft{\@ifnextchar^{\t@@left}{\t@@left^{}}}
 
\def\two@^#1_#2{\allowbreak
\def\n@one{#1}\def\n@two{#2}\mathrel{\setw@dth{#1}{#2}
\mathop{\vcenter{\lineskip\z@\baselineskip\z@
                 \hbox to \w@dth{\rightarrowfill}%
                 \hbox to \w@dth{\rightarrowfill}}%
       }\limits
\ifx\n@one\empty\else ^{\box\z@}\fi
\ifx\n@two\empty\else _{\box\@ne}\fi}}
\def\tw@@^#1{\@ifnextchar _{\two@^{#1}}{\two@^{#1}_{}}}
\def\two{\@ifnextchar ^{\tw@@}{\tw@@^{}}}
 
\def\tofr@^#1_#2{\def\n@one{#1}\def\n@two{#2}\mathrel{\setw@dth{#1}{#2}
\mathop{\vcenter{\hbox to \w@dth{\rightarrowfill}\kern-1.7ex
                 \hbox to \w@dth{\leftarrowfill}}%
       }\limits
\ifx\n@one\empty\else ^{\box\z@}\fi
\ifx\n@two\empty\else _{\box\@ne}\fi}}
\def\t@fr@^#1{\@ifnextchar_ {\tofr@^{#1}}{\tofr@^{#1}_{}}}
\def\tofro{\@ifnextchar^ {\t@fr@}{\t@fr@^{}}}

\def\mon{\mathop{\m@th\hbox to
      14.6\P@{\lasyb\char'51\hskip-2.1\P@$\arrext$\hss
$\mathord\rightarrow$}}\limits} 
\def\leftmono{\mathrel{\m@th\hbox to
14.6\P@{$\mathord\leftarrow$\hss$\arrext$\hskip-2.1\P@\lasyb\char'50%
}}\limits} 
\mathchardef\arrext="0200       

\def\settypes(#1,#2,#3){\arrowtypea#1 \arrowtypeb#2 \arrowtypec#3}
\def\settoheight#1#2{\setbox\@tempboxa\hbox{#2}#1\ht\@tempboxa\relax}%
\def\settodepth#1#2{\setbox\@tempboxa\hbox{#2}#1\dp\@tempboxa\relax}%
\def\settokens`#1`#2`#3`#4`{%
     \def\tokena{#1}\def\tokenb{#2}\def\tokenc{#3}\def\tokend{#4}}
\def\setsqparms[#1`#2`#3`#4;#5`#6]{%
\arrowtypea #1
\arrowtypeb #2
\arrowtypec #3
\arrowtyped #4
\width #5
\height #6
}
\def\setpos(#1,#2){\xpos=#1 \ypos#2}

\def\settriparms[#1`#2`#3;#4]{\settripairparms[#1`#2`#3`1`1;#4]}%
 
\def\settripairparms[#1`#2`#3`#4`#5;#6]{%
\arrowtypea #1
\arrowtypeb #2
\arrowtypec #3
\arrowtyped #4
\arrowtypee #5
\width #6
\height #6
}
 
\def\resetparms{\settripairparms[1`1`1`1`1;500]\width 500}
 
\resetparms
 
\def\mvector(#1,#2)#3{
\put(0,0){\vector(#1,#2){#3}}%
\put(0,0){\vector(#1,#2){26}}%
}
\def\evector(#1,#2)#3{{
\arrowlength #3
\put(0,0){\vector(#1,#2){\arrowlength}}%
\advance \arrowlength by-30
\put(0,0){\vector(#1,#2){\arrowlength}}%
}}
 
\def\horsize#1#2{%
\settowidth{\tempdimen}{$#2$}%
#1=\tempdimen
\divide #1 by\unitlength
}
 
\def\vertsize#1#2{%
\settoheight{\tempdimen}{$#2$}%
#1=\tempdimen
\settodepth{\tempdimen}{$#2$}%
\advance #1 by\tempdimen
\divide #1 by\unitlength
}
 
\def\putvector(#1,#2)(#3,#4)#5#6{{%
\ifnum3<\arrowtype
\putdashvector(#1,#2)(#3,#4)#5\arrowtype
\else
\ifnum\arrowtype<-3
\putdashvector(#1,#2)(#3,#4)#5\arrowtype
\else
\xpos=#1
\ypos=#2
\run=#3
\rise=#4
\arrowlength=#5
\ifnum \arrowtype<0
    \ifnum \run=0
        \advance \ypos by-\arrowlength
    \else
        \tempcounta \arrowlength
        \multiply \tempcounta by\rise
        \divide \tempcounta by\run
        \ifnum\run>0
            \advance \xpos by\arrowlength
            \advance \ypos by\tempcounta
        \else
            \advance \xpos by-\arrowlength
            \advance \ypos by-\tempcounta
        \fi
    \fi
    \multiply \arrowtype by-1
    \multiply \rise by-1
    \multiply \run by-1
\fi
\ifcase \arrowtype
\or \put(\xpos,\ypos){\vector(\run,\rise){\arrowlength}}%
\or \put(\xpos,\ypos){\mvector(\run,\rise)\arrowlength}%
\or \put(\xpos,\ypos){\evector(\run,\rise){\arrowlength}}%
\fi\fi\fi
}}
 
\def\putsplitvector(#1,#2)#3#4{
\xpos #1
\ypos #2
\arrowtype #4
\halflength #3
\arrowlength #3
\gap 140
\advance \halflength by-\gap
\divide \halflength by2
\ifnum\arrowtype>0
   \ifcase \arrowtype
   \or \put(\xpos,\ypos){\line(0,-1){\halflength}}%
       \advance\ypos by-\halflength
       \advance\ypos by-\gap
       \put(\xpos,\ypos){\vector(0,-1){\halflength}}%
   \or \put(\xpos,\ypos){\line(0,-1)\halflength}%
       \put(\xpos,\ypos){\vector(0,-1)3}%
       \advance\ypos by-\halflength
       \advance\ypos by-\gap
       \put(\xpos,\ypos){\vector(0,-1){\halflength}}%
   \or \put(\xpos,\ypos){\line(0,-1)\halflength}%
       \advance\ypos by-\halflength
       \advance\ypos by-\gap
       \put(\xpos,\ypos){\evector(0,-1){\halflength}}%
   \fi
\else \arrowtype=-\arrowtype
   \ifcase\arrowtype
   \or \advance \ypos by-\arrowlength
       \put(\xpos,\ypos){\line(0,1){\halflength}}%
       \advance\ypos by\halflength
       \advance\ypos by\gap
       \put(\xpos,\ypos){\vector(0,1){\halflength}}%
   \or \advance \ypos by-\arrowlength
       \put(\xpos,\ypos){\line(0,1)\halflength}%
       \put(\xpos,\ypos){\vector(0,1)3}%
       \advance\ypos by\halflength
       \advance\ypos by\gap
       \put(\xpos,\ypos){\vector(0,1){\halflength}}%
   \or \advance \ypos by-\arrowlength
       \put(\xpos,\ypos){\line(0,1)\halflength}%
       \advance\ypos by\halflength
       \advance\ypos by\gap
       \put(\xpos,\ypos){\evector(0,1){\halflength}}%
   \fi
\fi
}
 
\def\putmorphism(#1)(#2,#3)[#4`#5`#6]#7#8#9{{%
\run #2
\rise #3
\ifnum\rise=0
  \puthmorphism(#1)[#4`#5`#6]{#7}{#8}#9%
\else\ifnum\run=0
  \putvmorphism(#1)[#4`#5`#6]{#7}{#8}#9%
\else
\setpos(#1)%
\arrowlength #7
\arrowtype #8
\ifnum\run=0
\else\ifnum\rise=0
\else
\ifnum\run>0
    \coefa=1
\else
   \coefa=-1
\fi
\ifnum\arrowtype>0
   \coefb=0
   \coefc=-1
\else
   \coefb=\coefa
   \coefc=1
   \arrowtype=-\arrowtype
\fi
\width=2
\multiply \width by\run
\divide \width by\rise
\ifnum \width<0  \width=-\width\fi
\advance\width by60
\if l#9 \width=-\width\fi
\putbox(\xpos,\ypos){#4}
{\multiply \coefa by\arrowlength
\advance\xpos by\coefa
\multiply \coefa by\rise
\divide \coefa by\run
\advance \ypos by\coefa
\putbox(\xpos,\ypos){#5} }%
{\multiply \coefa by\arrowlength
\divide \coefa by2
\advance \xpos by\coefa
\advance \xpos by\width
\multiply \coefa by\rise
\divide \coefa by\run
\advance \ypos by\coefa
\if l#9%
   \putrbox(\xpos,\ypos){#6}%
\else\if r#9%
   \putlbox(\xpos,\ypos){#6}%
\fi\fi }%
{\multiply \rise by-\coefc
\multiply \run by-\coefc
\multiply \coefb by\arrowlength
\advance \xpos by\coefb
\multiply \coefb by\rise
\divide \coefb by\run
\advance \ypos by\coefb
\multiply \coefc by70
\advance \ypos by\coefc
\multiply \coefc by\run
\divide \coefc by\rise
\advance \xpos by\coefc
\multiply \coefa by140
\multiply \coefa by\run
\divide \coefa by\rise
\advance \arrowlength by\coefa
\ifcase\arrowtype
\or \put(\xpos,\ypos){\vector(\run,\rise){\arrowlength}}%
\or \put(\xpos,\ypos){\mvector(\run,\rise){\arrowlength}}%
\or \put(\xpos,\ypos){\evector(\run,\rise){\arrowlength}}%
\fi}\fi\fi\fi\fi}}

\newcount\numbdashes \newcount\lengthdash \newcount\increment
 
\def\howmanydashes{
\numbdashes=\arrowlength \lengthdash=40
\divide\numbdashes by \lengthdash
\lengthdash=\arrowlength
\divide\lengthdash by \numbdashes
\increment=\lengthdash
\multiply\lengthdash by 3
\divide\lengthdash by 5
}
 
\def\putdashvector(#1)(#2,#3)#4#5{%
\ifnum#3=0 \putdashhvector(#1){#4}#5
\else
\ifnum#2=0
\putdashvvector(#1){#4}#5\fi\fi}
 
\def\putdashhvector(#1,#2)#3#4{{%
\arrowlength=#3 \howmanydashes
\multiput(#1,#2)(\increment,0){\numbdashes}%
{\vrule height .4pt width \lengthdash\unitlength}
\arrowtype=#4 \xpos=#1
\ifnum\arrowtype<0 \advance\arrowtype by 7 \fi
\ifcase\arrowtype
\or \advance\xpos by 10
    \put(\xpos,#2){\vector(-1,0){\lengthdash}}
    \advance\xpos by 40
    \put(\xpos,#2){\vector(-1,0){\lengthdash}}
\or \advance \xpos by 10
    \put(\xpos,#2){\vector(-1,0){\lengthdash}}
    \advance\xpos by  \arrowlength
    \advance\xpos by  -50
    \put(\xpos,#2){\vector(-1,0){\lengthdash}}
\or \advance\xpos by 10
    \put(\xpos,#2){\vector(-1,0){\lengthdash}}
\or \advance\xpos by \arrowlength
    \advance\xpos by -\lengthdash
    \put(\xpos,#2){\vector(1,0){\lengthdash}}
\or {\advance\xpos by 10
    \put(\xpos,#2){\vector(1,0){\lengthdash}}}
    \advance\xpos by \arrowlength
    \advance\xpos by -\lengthdash
    \put(\xpos,#2){\vector(1,0){\lengthdash}}
\or \advance\xpos by \arrowlength
    \advance\xpos by -\lengthdash
    \put(\xpos,#2){\vector(1,0){\lengthdash}}
    \advance\xpos by -40
    \put(\xpos,#2){\vector(1,0){\lengthdash}}
   \fi
}}
 
\def\putdashvvector(#1,#2)#3#4{{%
\arrowlength=#3 \howmanydashes
\ypos=#2 \advance\ypos by -\arrowlength
\multiput(#1,#2)(0,\increment){\numbdashes}%
    {\vrule width .4pt height \lengthdash\unitlength}
\arrowtype=#4 \ypos=#2
\ifnum\arrowtype<0 \advance\arrowtype by 7 \fi
\ifcase\arrowtype
\or \advance\ypos by \arrowlength \advance\ypos by -40
    \put(#1,\ypos){\vector(0,1){\lengthdash}}
    \advance\ypos by -40
    \put(#1,\ypos){\vector(0,1){\lengthdash}}
\or \advance\ypos by 10
    \put(#1,\ypos){\vector(0,1){\lengthdash}}
    \advance\ypos by \arrowlength \advance\ypos by -40
    \put(#1,\ypos){\vector(0,1){\lengthdash}}
\or \advance\ypos by \arrowlength \advance\ypos by -40
    \put(#1,\ypos){\vector(0,1){\lengthdash}}
\or \advance\ypos by 10
    \put(#1,\ypos){\vector(0,-1){\lengthdash}}
\or \advance\ypos by 10
    \put(#1,\ypos){\vector(0,-1){\lengthdash}}
    \advance\ypos by \arrowlength \advance\ypos by -40
    \put(#1,\ypos){\vector(0,-1){\lengthdash}}
\or \advance\ypos by 10
    \put(#1,\ypos){\vector(0,-1){\lengthdash}}
    \advance\ypos by 40
    \put(#1,\ypos){\vector(0,-1){\lengthdash}}
\fi
}}
 
\def\puthmorphism(#1,#2)[#3`#4`#5]#6#7#8{{%
\xpos #1
\ypos #2
\width #6
\arrowlength #6
\arrowtype=#7
\putbox(\xpos,\ypos){#3\vphantom{#4}}%
{\advance \xpos by\arrowlength
\putbox(\xpos,\ypos){\vphantom{#3}#4}}%
\horsize{\tempcounta}{#3}%
\horsize{\tempcountb}{#4}%
\divide \tempcounta by2
\divide \tempcountb by2
\advance \tempcounta by30
\advance \tempcountb by30
\advance \xpos by\tempcounta
\advance \arrowlength by-\tempcounta
\advance \arrowlength by-\tempcountb
\putvector(\xpos,\ypos)(1,0)\arrowlength\arrowtype
\divide \arrowlength by2
\advance \xpos by\arrowlength
\vertsize{\tempcounta}{#5}%
\divide\tempcounta by2
\advance \tempcounta by20
\if a#8 %
   \advance \ypos by\tempcounta
   \putbox(\xpos,\ypos){#5}%
\else
   \advance \ypos by-\tempcounta
   \putbox(\xpos,\ypos){#5}%
\fi}}
 
\def\putvmorphism(#1,#2)[#3`#4`#5]#6#7#8{{%
\xpos #1
\ypos #2
\arrowlength #6
\arrowtype #7
\settowidth{\xlen}{$#5$}%
\putbox(\xpos,\ypos){#3}%
{\advance \ypos by-\arrowlength
\putbox(\xpos,\ypos){#4}}%
{\advance\arrowlength by-140
\advance \ypos by-70
\ifdim\xlen>0pt
   \if m#8%
      \putsplitvector(\xpos,\ypos)\arrowlength\arrowtype
   \else
   \putvector(\xpos,\ypos)(0,-1)\arrowlength\arrowtype
   \fi
\else
   \putvector(\xpos,\ypos)(0,-1)\arrowlength\arrowtype
\fi}%
\ifdim\xlen>0pt
   \divide \arrowlength by2
   \advance\ypos by-\arrowlength
   \if l#8%
      \advance \xpos by-40
      \putrbox(\xpos,\ypos){#5}%
   \else\if r#8%
      \advance \xpos by40
      \putlbox(\xpos,\ypos){#5}%
   \else
      \putbox(\xpos,\ypos){#5}%
   \fi\fi
\fi
}}
 
\def\putsqquarep<#1>(#2)[#3;#4`#5`#6`#7]{{%
\setsqparms[#1]%
\setpos(#2)%
\settokens`#3`%
\puthmorphism(\xpos,\ypos)[\tokenc`\tokend`{#7}]{\width}{\arrowtyped}b%
\advance\ypos by \height
\puthmorphism(\xpos,\ypos)[\tokena`\tokenb`{#4}]{\width}{\arrowtypea}a%
\putvmorphism(\xpos,\ypos)[``{#5}]{\height}{\arrowtypeb}l%
\advance\xpos by \width
\putvmorphism(\xpos,\ypos)[``{#6}]{\height}{\arrowtypec}r%
}}
 
\def\putsqquare{\@ifnextchar <{\putsqquarep}{\putsqquarep%
   <\arrowtypea`\arrowtypeb`\arrowtypec`\arrowtyped;\width`\height>}}
\def\sqquare{\@ifnextchar< {\sqquarep}{\sqquarep
   <\arrowtypea`\arrowtypeb`\arrowtypec`\arrowtyped;\width`\height>}}
\def\sqquarep<#1>[#2`#3`#4`#5;#6`#7`#8`#9]{{
\setsqparms[#1]
\diagram
\putsqquarep<\arrowtypea`\arrowtypeb`\arrowtypec`
\arrowtyped;\width`\height>
(0,0)[#2`#3`#4`{#5};#6`#7`#8`{#9}]
\enddiagram
}}                                                 
\def\putptrianglep<#1>(#2,#3)[#4`#5`#6;#7`#8`#9]{{%
\settriparms[#1]%
\xpos=#2 \ypos=#3
\advance\ypos by \height
\puthmorphism(\xpos,\ypos)[#4`#5`{#7}]{\height}{\arrowtypea}a%
\putvmorphism(\xpos,\ypos)[`#6`{#8}]{\height}{\arrowtypeb}l%
\advance\xpos by\height
\putmorphism(\xpos,\ypos)(-1,-1)[``{#9}]{\height}{\arrowtypec}r%
}}
 
\def\putptriangle{\@ifnextchar <{\putptrianglep}{\putptrianglep
   <\arrowtypea`\arrowtypeb`\arrowtypec;\height>}}
\def\ptriangle{\@ifnextchar <{\ptrianglep}{\ptrianglep
   <\arrowtypea`\arrowtypeb`\arrowtypec;\height>}}
\def\ptrianglep<#1>[#2`#3`#4;#5`#6`#7]{{
\settriparms[#1]
\diagram
\putptrianglep<\arrowtypea`\arrowtypeb`
\arrowtypec;\height>
(0,0)[#2`#3`#4;#5`#6`{#7}]
\enddiagram
}}                                            
 
\def\putqtrianglep<#1>(#2,#3)[#4`#5`#6;#7`#8`#9]{{%
\settriparms[#1]%
\xpos=#2 \ypos=#3
\advance\ypos by\height
\puthmorphism(\xpos,\ypos)[#4`#5`{#7}]{\height}{\arrowtypea}a%
\putmorphism(\xpos,\ypos)(1,-1)[``{#8}]{\height}{\arrowtypeb}l%
\advance\xpos by\height
\putvmorphism(\xpos,\ypos)[`#6`{#9}]{\height}{\arrowtypec}r%
}}
 
\def\putqtriangle{\@ifnextchar <{\putqtrianglep}{\putqtrianglep
   <\arrowtypea`\arrowtypeb`\arrowtypec;\height>}}
\def\qtriangle{\@ifnextchar <{\qtrianglep}{\qtrianglep
   <\arrowtypea`\arrowtypeb`\arrowtypec;\height>}}
\def\qtrianglep<#1>[#2`#3`#4;#5`#6`#7]{{
\settriparms[#1]
\width=\height                                
\diagram
\putqtrianglep<\arrowtypea`\arrowtypeb`
\arrowtypec;\height>
(0,0)[#2`#3`#4;#5`#6`{#7}]
\enddiagram
}}
 
\def\putdtrianglep<#1>(#2,#3)[#4`#5`#6;#7`#8`#9]{{%
\settriparms[#1]%
\xpos=#2 \ypos=#3
\puthmorphism(\xpos,\ypos)[#5`#6`{#9}]{\height}{\arrowtypec}b%
\advance\xpos by \height \advance\ypos by\height
\putmorphism(\xpos,\ypos)(-1,-1)[``{#7}]{\height}{\arrowtypea}l%
\putvmorphism(\xpos,\ypos)[#4``{#8}]{\height}{\arrowtypeb}r%
}}
 
\def\putdtriangle{\@ifnextchar <{\putdtrianglep}{\putdtrianglep
   <\arrowtypea`\arrowtypeb`\arrowtypec;\height>}}
\def\dtriangle{\@ifnextchar <{\dtrianglep}{\dtrianglep
   <\arrowtypea`\arrowtypeb`\arrowtypec;\height>}}
\def\dtrianglep<#1>[#2`#3`#4;#5`#6`#7]{{
\settriparms[#1]
\width=\height                                
\diagram
\putdtrianglep<\arrowtypea`\arrowtypeb`
\arrowtypec;\height>
(0,0)[#2`#3`#4;#5`#6`{#7}]
\enddiagram
}}
 
\def\putbtrianglep<#1>(#2,#3)[#4`#5`#6;#7`#8`#9]{{%
\settriparms[#1]%
\xpos=#2 \ypos=#3
\puthmorphism(\xpos,\ypos)[#5`#6`{#9}]{\height}{\arrowtypec}b%
\advance\ypos by\height
\putmorphism(\xpos,\ypos)(1,-1)[``{#8}]{\height}{\arrowtypeb}r%
\putvmorphism(\xpos,\ypos)[#4``{#7}]{\height}{\arrowtypea}l%
}}
 
\def\putbtriangle{\@ifnextchar <{\putbtrianglep}{\putbtrianglep
   <\arrowtypea`\arrowtypeb`\arrowtypec;\height>}}
\def\btriangle{\@ifnextchar <{\btrianglep}{\btrianglep
   <\arrowtypea`\arrowtypeb`\arrowtypec;\height>}}
\def\btrianglep<#1>[#2`#3`#4;#5`#6`#7]{{
\settriparms[#1]
\width=\height                               
\diagram
\putbtrianglep<\arrowtypea`\arrowtypeb`
\arrowtypec;\height>
(0,0)[#2`#3`#4;#5`#6`{#7}]
\enddiagram
}}
 
\def\putAtrianglep<#1>(#2,#3)[#4`#5`#6;#7`#8`#9]{{%
\settriparms[#1]%
\xpos=#2 \ypos=#3
{\multiply \height by2
\puthmorphism(\xpos,\ypos)[#5`#6`{#9}]{\height}{\arrowtypec}b}%
\advance\xpos by\height \advance\ypos by\height
\putmorphism(\xpos,\ypos)(-1,-1)[#4``{#7}]{\height}{\arrowtypea}l%
\putmorphism(\xpos,\ypos)(1,-1)[``{#8}]{\height}{\arrowtypeb}r%
}}
 
\def\putAtriangle{\@ifnextchar <{\putAtrianglep}{\putAtrianglep
   <\arrowtypea`\arrowtypeb`\arrowtypec;\height>}}
\def\Atriangle{\@ifnextchar <{\Atrianglep}{\Atrianglep
   <\arrowtypea`\arrowtypeb`\arrowtypec;\height>}}
\def\Atrianglep<#1>[#2`#3`#4;#5`#6`#7]{{
\settriparms[#1]
\width=\height                                     
\diagram
\putAtrianglep<\arrowtypea`\arrowtypeb`
\arrowtypec;\height>
(0,0)[#2`#3`#4;#5`#6`{#7}]
\enddiagram
}}
 
\def\putAtrianglepairp<#1>(#2)[#3;#4`#5`#6`#7`#8]{{%
\settripairparms[#1]%
\setpos(#2)%
\settokens`#3`%
\puthmorphism(\xpos,\ypos)[\tokenb`\tokenc`{#7}]{\height}{\arrowtyped}b%
\advance\xpos by\height
\puthmorphism(\xpos,\ypos)[\phantom{\tokenc}`\tokend`{#8}]%
{\height}{\arrowtypee}b%
\advance\ypos by\height
\putmorphism(\xpos,\ypos)(-1,-1)[\tokena``{#4}]{\height}{\arrowtypea}l%
\putvmorphism(\xpos,\ypos)[``{#5}]{\height}{\arrowtypeb}m%
\putmorphism(\xpos,\ypos)(1,-1)[``{#6}]{\height}{\arrowtypec}r%
}}
 
\def\putAtrianglepair{\@ifnextchar <{\putAtrianglepairp}{\putAtrianglepairp%
   <\arrowtypea`\arrowtypeb`\arrowtypec`\arrowtyped`\arrowtypee;\height>}}
\def\Atrianglepair{\@ifnextchar <{\Atrianglepairp}{\Atrianglepairp%
   <\arrowtypea`\arrowtypeb`\arrowtypec`\arrowtyped`\arrowtypee;\height>}}
 
\def\Atrianglepairp<#1>[#2;#3`#4`#5`#6`#7]{{
\settripairparms[#1]
\settokens`#2`
\width=\height                                
\diagram
\putAtrianglepairp                            
<\arrowtypea`\arrowtypeb`\arrowtypec`
\arrowtyped`\arrowtypee;\height>
(0,0)[{#2};#3`#4`#5`#6`{#7}]
\enddiagram
}}
 
\def\putVtrianglep<#1>(#2,#3)[#4`#5`#6;#7`#8`#9]{{%
\settriparms[#1]%
\xpos=#2 \ypos=#3
\advance\ypos by\height
{\multiply\height by2
\puthmorphism(\xpos,\ypos)[#4`#5`{#7}]{\height}{\arrowtypea}a}%
\putmorphism(\xpos,\ypos)(1,-1)[`#6`{#8}]{\height}{\arrowtypeb}l%
\advance\xpos by\height
\advance\xpos by\height
\putmorphism(\xpos,\ypos)(-1,-1)[``{#9}]{\height}{\arrowtypec}r%
}}
 
\def\putVtriangle{\@ifnextchar <{\putVtrianglep}{\putVtrianglep
   <\arrowtypea`\arrowtypeb`\arrowtypec;\height>}}
\def\Vtriangle{\@ifnextchar <{\Vtrianglep}{\Vtrianglep
   <\arrowtypea`\arrowtypeb`\arrowtypec;\height>}}
\def\Vtrianglep<#1>[#2`#3`#4;#5`#6`#7]{{
\settriparms[#1]
\width=\height                                 
\diagram
\putVtrianglep<\arrowtypea`\arrowtypeb`
\arrowtypec;\height>
(0,0)[#2`#3`#4;#5`#6`{#7}]
\enddiagram
}}
 
\def\putVtrianglepairp<#1>(#2)[#3;#4`#5`#6`#7`#8]{{
\settripairparms[#1]%
\setpos(#2)%
\settokens`#3`%
\advance\ypos by\height
\putmorphism(\xpos,\ypos)(1,-1)[`\tokend`{#6}]{\height}{\arrowtypec}l%
\puthmorphism(\xpos,\ypos)[\tokena`\tokenb`{#4}]{\height}{\arrowtypea}a%
\advance\xpos by\height
\puthmorphism(\xpos,\ypos)[\phantom{\tokenb}`\tokenc`{#5}]%
{\height}{\arrowtypeb}a%
\putvmorphism(\xpos,\ypos)[``{#7}]{\height}{\arrowtyped}m%
\advance\xpos by\height
\putmorphism(\xpos,\ypos)(-1,-1)[``{#8}]{\height}{\arrowtypee}r%
}}
 
\def\putVtrianglepair{\@ifnextchar <{\putVtrianglepairp}{\putVtrianglepairp%
    <\arrowtypea`\arrowtypeb`\arrowtypec`\arrowtyped`\arrowtypee;\height>}}
\def\Vtrianglepair{\@ifnextchar <{\Vtrianglepairp}{\Vtrianglepairp%
    <\arrowtypea`\arrowtypeb`\arrowtypec`\arrowtyped`\arrowtypee;\height>}}
\def\Vtrianglepairp<#1>[#2;#3`#4`#5`#6`#7]{{
\settripairparms[#1]
\settokens`#2`
\diagram
\putVtrianglepairp                             
<\arrowtypea`\arrowtypeb`\arrowtypec`
\arrowtyped`\arrowtypee;\height>
(0,0)[{#2};#3`#4`#5`#6`{#7}]
\enddiagram
}}

\def\putCtrianglep<#1>(#2,#3)[#4`#5`#6;#7`#8`#9]{{%
\settriparms[#1]%
\xpos=#2 \ypos=#3
\advance\ypos by\height
\putmorphism(\xpos,\ypos)(1,-1)[``{#9}]{\height}{\arrowtypec}l%
\advance\xpos by\height
\advance\ypos by\height
\putmorphism(\xpos,\ypos)(-1,-1)[#4`#5`{#7}]{\height}{\arrowtypea}l%
{\multiply\height by 2
\putvmorphism(\xpos,\ypos)[`#6`{#8}]{\height}{\arrowtypeb}r}%
}}
 
\def\putCtriangle{\@ifnextchar <{\putCtrianglep}{\putCtrianglep
    <\arrowtypea`\arrowtypeb`\arrowtypec;\height>}}
\def\Ctriangle{\@ifnextchar <{\Ctrianglep}{\Ctrianglep
    <\arrowtypea`\arrowtypeb`\arrowtypec;\height>}}
\def\Ctrianglep<#1>[#2`#3`#4;#5`#6`#7]{{
\settriparms[#1]
\width=\height                               
\diagram
\putCtrianglep<\arrowtypea`\arrowtypeb`
\arrowtypec;\height>
(0,0)[#2`#3`#4;#5`#6`{#7}]
\enddiagram
}}                                           
\def\putDtrianglep<#1>(#2,#3)[#4`#5`#6;#7`#8`#9]{{%
\settriparms[#1]%
\xpos=#2 \ypos=#3
\advance\xpos by\height \advance\ypos by\height
\putmorphism(\xpos,\ypos)(-1,-1)[``{#9}]{\height}{\arrowtypec}r%
\advance\xpos by-\height \advance\ypos by\height
\putmorphism(\xpos,\ypos)(1,-1)[`#5`{#8}]{\height}{\arrowtypeb}r%
{\multiply\height by 2
\putvmorphism(\xpos,\ypos)[#4`#6`{#7}]{\height}{\arrowtypea}l}%
}}
 
\def\putDtriangle{\@ifnextchar <{\putDtrianglep}{\putDtrianglep
    <\arrowtypea`\arrowtypeb`\arrowtypec;\height>}}
\def\Dtriangle{\@ifnextchar <{\Dtrianglep}{\Dtrianglep
   <\arrowtypea`\arrowtypeb`\arrowtypec;\height>}}
\def\Dtrianglep<#1>[#2`#3`#4;#5`#6`#7]{{
\settriparms[#1]
\width=\height                              
\diagram
\putDtrianglep<\arrowtypea`\arrowtypeb`
\arrowtypec;\height>
(0,0)[#2`#3`#4;#5`#6`{#7}]
\enddiagram
}}                                          
\def\setrecparms[#1`#2]{\width=#1 \height=#2}%
 
\def\recursep<#1`#2>[#3;#4`#5`#6`#7`#8]{{\m@th
\width=#1 \height=#2
\settokens`#3`
\settowidth{\tempdimen}{$\tokena$}
\ifdim\tempdimen=0pt
  \savebox{\tempboxa}{\hbox{$\tokenb$}}%
  \savebox{\tempboxb}{\hbox{$\tokend$}}%
  \savebox{\tempboxc}{\hbox{$#6$}}%
\else
  \savebox{\tempboxa}{\hbox{$\hbox{$\tokena$}\times\hbox{$\tokenb$}$}}%
  \savebox{\tempboxb}{\hbox{$\hbox{$\tokena$}\times\hbox{$\tokend$}$}}%
  \savebox{\tempboxc}{\hbox{$\hbox{$\tokena$}\times\hbox{$#6$}$}}%
\fi
\ypos=\height
\divide\ypos by 2
\xpos=\ypos
\advance\xpos by \width
\bfig
\putCtrianglep<-1`1`1;\ypos>(0,0)[`\tokenc`;#5`#6`{#7}]%
\puthmorphism(\ypos,0)[\tokend`\usebox{\tempboxb}`{#8}]{\width}{-1}b%
\puthmorphism(\ypos,\height)[\tokenb`\usebox{\tempboxa}`{#4}]{\width}{-1}a%
\advance\ypos by \width
\putvmorphism(\ypos,\height)[``\usebox{\tempboxc}]{\height}1r%
\efig
}}
 
\def\recurse{\@ifnextchar <{\recursep}{\recursep<\width`\height>}}
 
\def\puttwohmorphisms(#1,#2)[#3`#4;#5`#6]#7#8#9{{%
\puthmorphism(#1,#2)[#3`#4`]{#7}0a
\ypos=#2
\advance\ypos by 20
\puthmorphism(#1,\ypos)[\phantom{#3}`\phantom{#4}`#5]{#7}{#8}a
\advance\ypos by -40
\puthmorphism(#1,\ypos)[\phantom{#3}`\phantom{#4}`#6]{#7}{#9}b
}}
 
\def\puttwovmorphisms(#1,#2)[#3`#4;#5`#6]#7#8#9{{%
\putvmorphism(#1,#2)[#3`#4`]{#7}0a
\xpos=#1
\advance\xpos by -20
\putvmorphism(\xpos,#2)[\phantom{#3}`\phantom{#4}`#5]{#7}{#8}l
\advance\xpos by 40
\putvmorphism(\xpos,#2)[\phantom{#3}`\phantom{#4}`#6]{#7}{#9}r
}}
 
\def\puthcoequalizer(#1)[#2`#3`#4;#5`#6`#7]#8#9{{%
\setpos(#1)%
\puttwohmorphisms(\xpos,\ypos)[#2`#3;#5`#6]{#8}11%
\advance\xpos by #8
\puthmorphism(\xpos,\ypos)[\phantom{#3}`#4`#7]{#8}1{#9}
}}
 
\def\putvcoequalizer(#1)[#2`#3`#4;#5`#6`#7]#8#9{{%
\setpos(#1)%
\puttwovmorphisms(\xpos,\ypos)[#2`#3;#5`#6]{#8}11%
\advance\ypos by -#8
\putvmorphism(\xpos,\ypos)[\phantom{#3}`#4`#7]{#8}1{#9}
}}
 
\def\putthreehmorphisms(#1)[#2`#3;#4`#5`#6]#7(#8)#9{{%
\setpos(#1) \settypes(#8)
\if a#9 %
     \vertsize{\tempcounta}{#5}%
     \vertsize{\tempcountb}{#6}%
     \ifnum \tempcounta<\tempcountb \tempcounta=\tempcountb \fi
\else
     \vertsize{\tempcounta}{#4}%
     \vertsize{\tempcountb}{#5}%
     \ifnum \tempcounta<\tempcountb \tempcounta=\tempcountb \fi
\fi
\advance \tempcounta by 60
\puthmorphism(\xpos,\ypos)[#2`#3`#5]{#7}{\arrowtypeb}{#9}
\advance\ypos by \tempcounta
\puthmorphism(\xpos,\ypos)[\phantom{#2}`\phantom{#3}`#4]{#7}{\arrowtypea}{#9}
\advance\ypos by -\tempcounta \advance\ypos by -\tempcounta
\puthmorphism(\xpos,\ypos)[\phantom{#2}`\phantom{#3}`#6]{#7}{\arrowtypec}{#9}
}}
 
\def\setarrowtoks[#1`#2`#3`#4`#5`#6]{%
\def\toka{#1}
\def\tokb{#2}
\def\tokc{#3}
\def\tokd{#4}
\def\toke{#5}
\def\tokf{#6}
}
\def\hex{\@ifnextchar <{\hexp}{\hexp<1000`400>}}
\def\hexp<#1`#2>[#3`#4`#5`#6`#7`#8;#9]{%
\setarrowtoks[#9]
\yext=#2 \advance \yext by #2
\xext=#1 \advance\xext by \yext
\bfig
\putCtriangle<-1`0`1;#2>(0,0)[`#5`;\tokb``\tokd]
\xext=#1 \yext=#2 \advance \yext by #2
\putsqquare<1`0`0`1;\xext`\yext>(#2,0)[#3`#4`#7`#8;\toka```\tokf]
\advance \xext by #2
\putDtriangle<0`1`-1;#2>(\xext,0)[`#6`;`\tokc`\toke]
\efig
}
\makeatother

\usepackage{./axodraw}
\begin{document}
\def\mkb{\mbox}
\def\beq{\begin{equation}}
\def\eeq{\end{equation}}
\def\beqn{\begin{eqnarray}}
\def\eeqn{\end{eqnarray}}
\def\alg{\mathcal{A}}
\def\vNalg{\mathcal{N}}
\def\intv{\mathcal{I}}
\def\W{\mathcal{W}}
\def\R{\mathbb{R}}
\def\C{\mathbb{C}}
\def\Z{\mathbb{Z}}
\def\N{\mathbb{N}}
\thispagestyle{empty}
\begin{flushright}
                                       BONN--TH--2000--11\\
\end{flushright}
\vspace{1cm}
\begin{center}
     {\Large{\bf Comment on: Modular Theory and Geometry}}\\
\vspace{2cm}
    K. Ebrahimi-Fard
\footnote{
          e-mail: fard@th.physik.uni-bonn.de
         }\\
\vspace{1.5cm}
\vspace{0.2cm}
{\sl Physikalisches Institut der Universit\"at Bonn}\\
{\sl Nu{\ss}allee 12, D--53115 Bonn, Germany}
\end{center}
\vspace{2.5cm}
\setcounter{page}{1}
\begin{abstract}
In this note we comment on part of a recent article by B. Schroer and
H.-W. Wiesbrock. Therein they calculate some new modular structure for
the $U(1)$-current-algebra (Weyl-algebra). We point out that their
findings are true in a more general setting.
\end{abstract}
\newpage

{\Large{\bf{I}}}
We would like to add a point to a recent inspiring work of B. Schroer and
H.-W. Wiesbrock \cite{SW}\footnote{
Recently this work was corrected in an important point and also extended by
Schroer and Fassarella \cite{SF}. Their new results demanded also for a
correction of an earlier version of this paper.}
implying a modular origin of the chiral "higher dilation" diffeomorphisms.\\
For the $U(1)$-current-algebra in two-dimensional
spacetime they construct states invariant under higher representations of
the M\"obius-group, generated by the modes $L_{-n,0,n}$.
The new states fulfill the KMS-property with respect to modified dilations.
These modified dilations are \mkb{identified} with the modular group associated
to von Neumann algebras localized in well-chosen regions and the new states.\\
In the original version \cite{SW} of the conjecture it was overlooked that the
chosen state lacked the property of faithfulness if used on doubly-localized
intervals.
As far as the demonstration of the modular origin of the diffeomorphism group was
concerned this was corrected in the work of Schroer and Fassarella \cite{SF} but the
correction was at the expense of the original conjecture.
Here we show that due to the geometrical properties of the modified transformations
respectively the new F-S states the use of the split-property \cite{Haag} allows a
faithful extension of these F-S states to doubly-localized intervals and in this way
the demonstration of the modular origin of the diffeomorphism group is harmonized
with the understanding of the modular structure of double intervals.\\
In section 2 we sketch briefly the ansatz and result
concerning the \mkb{modular} structure in the case of the
$U(1)$-current-algebra.
Section 3 contains our point to add, providing a more
general point of view of the aforementioned results of
Schroer and Wiesbrock.
In section 4 we investigate socalled multilocalized algebras
by using the split-property. The product states and modular
group of such algebras are identified.
Some additional remarks and a short summary are given in Section 5.\\[0.3cm]
%

{\Large{\bf{II}}}
Conformal field theory in two dimensions \mkb{($CFT_{2}$)} \cite{DiF}
provides a well suited realm for algebraic quantum field theory \cite{W1}
\footnote{
 For a very recent review of the current state of algebraic QFT see \cite{DB}.
          },
especially for problems concerning the geometric identification of the
modular \mkb{structure \cite{BW,B2}}.\\
Minkowskian \mkb{$CFT_{2}$} may be represented on the product of two
circles, \mkb{$S^{1} \times S^{1}$}-space\-time (the ''compact-
picture''). The global symmetry group of the \mkb{$CFT_{2}$} is the
M\"obius-group \mkb{$PSU(1,1) \times PSU(1,1)$}. We will concentrate on one of
the groups, being realized on one of the circles: \[
PSU(1,1):=SU(1,1)/\{ \pm1 \},
\]
\[
 SU(1,1):=\left\lbrace
     \left(\matrix{
                   \alpha & \beta \cr
              \bar{\beta} & \bar{\alpha} \cr
                   }\right) \Bigg|
                    \:\alpha,\beta \in \C,\; |\alpha|^{2}-|\beta|^{2}=1
           \right\rbrace.
\]
The spectrum generating algebra of reparametrizations of the circle
is \mkb{generated} by the Virasoro algebra (with central charge $c$)
\mkb{$\mathfrak{L}_{c}$}:
\beq
 [L_{n},L_{m}]=(n-m)L_{n+m}+\frac{c}{12}(n^{3}-n)\delta_{n+m,0}
 ,\;\;\; n \in \Z.
 \label{Vir}
\eeq
The globally realized group \mkb{$PSU(1,1)$} has the underlying generators
\mkb{$L_{-1},L_{0},L_{1}$}, fulfilling a $sl(2,\C)$-algebra:
\beq
[L_{1},L_{-1}]=2L_{0}, \;\;\; [L_{\pm 1},L_{0}]=\pm L_{\pm 1}.
\eeq
The Virasoro algebra \mkb{$\mathfrak{L}_{c}$} contains infinitely many
further $sl(2,\C)$-algebras, generated by the modes
\mkb{$L_{-n},L_{0},L_{n},\; n>1$}:
\beq
  \matrix{
  \!\!\!\!\!\!\!\!\!\!\!\!\!\!\!\!\!\!\!\!\!\!\!\!\!\!\!\!L_{-n}
  \mapsto \tilde{L}_{-n} := \frac{1}{n}L_{-n} \cr
  L_{0}  \mapsto \: \tilde{L}_{0} \;:= \frac{1}{n}L_{0} +
                                            \frac{c}{24}\frac{(n^{2}-1)}{n}\cr
  \!\!\!\!\!\!\!\!\!\!\!\!\!\!\!\!\!\!\!\!\!\!\!\!\!\!\!\!L_{+n}
  \mapsto \tilde{L}_{+n} := \frac{1}{n}L_{+n}
          }
  \;\;\;\longmapsto \Bigg\{
           \matrix{
           \![\tilde{L}_{+n},\tilde{L}_{-n}]=2\tilde{L}_{0}\cr
         \:[\tilde{L}_{\pm n},\tilde{L}_{0}]= \pm \tilde{L}_{\pm n}
                   }\Bigg\}\:sl(2,\C).
                                                       \label{nal}
\eeq
The corresponding finite transformations are of the form:
\beq
  g_{n}(z):= \left( \frac{\alpha z^{n} + \beta}
             {\bar{\beta}z^{n}+\bar{\alpha}} \right)^{\frac{1}{n}}\!\!,
          \;\;\;\left(\matrix{
                            \alpha & \beta \cr
                       \bar{\beta} & \bar{\alpha} \cr
                            }\right) \in  PSU(1,1).
                                                        \label{nnt}
\eeq
They leave the unit-circle $S^{1}$ invariant.\\
One may equally well represent the \mkb{$CFT_{2}$} on a product of lines,
\mkb{$\R \times \R$}-spacetime (the ''non-compact-picture'').
The coordinate transformation from the circle to the line is provided by the
stereographic projection (Cayley-transformation):
\begin{equation}
   S^{1}\backslash\{-1\} \ni z \longmapsto x(z):= -i \frac{z-1}{z+1} \in
\R,\;\; -1    \longmapsto \infty.
                                                        \label{cayley}
\end{equation}
The global symmetry group \mkb{$PSU(1,1)$} transforms in this
process isomorphically into the real group \mkb{$PSL(2,\R)$} \cite{FG}:
\[
 \R \ni x \longmapsto \hat{g}(x):= \frac{ax + b}{cx + d},\;\;\;ad-bc=1.
\]
It is slightly more cumbersome to handle the transformations analogue to
(eq. \ref{nnt}) in the non-compact picture. For this reason we perform the
calculations in the compact picture representation.\\
Schroer and Wiesbrock take as a paradigm of their discussion the
$U(1)$-current-algebra on the circle. The constituting relation of this
$U(1)$-algebra is the current-current commutation-relation (with the circle as
base-space):
$$
          [J(z),J(w)]=-\partial_{z}\delta(z-w).
$$
In order to bring the algebraic ansatz \cite{Haag} to the stage one has to
smear the currents by real testfunctions on the circle:
\[
  J(f):= \int_{S^{1}}\:\frac{dz}{2\pi i} f(z)\: J(z).
\]
The bounded operators \mkb{$e^{i J(f)}$} give rise to a von Neumann
algebra, i.e. weakly closed algebra of bounded operators:
\[
\W(\intv):=\left\{
            W(f):=e^{iJ(f)} \bigg| supp(f) \subseteq \intv \subset S^{1}
           \right\}'',
\]
which is called (local) Weyl-algebra \cite{BMT, BR}.
The double prime indicates here the double-commutant which after a
theorem by von Neumann (double-commu\-tant-theorem \cite{BR}) equals
the weak closure of the algebra generated by the \mkb{$e^{i J(f)}$}.
The net of Weyl-algebras
\mkb{$\{S^{1} \supset \intv \mapsto \W(\intv)\}$}
fulfills the postulates of algebraic quantum field theory, in
particular locality \cite{BS}.\\
The one-parameter group of dilations in \mkb{$PSU(1,1)$} has the following
form:
\beq
  Dil(t)z = \frac{ch(\pi t)z+sh(\pi t)}
                               {sh(\pi t)z+ch(\pi t)}
        ,\;\;\;t \in \R,\;z\in S^{1}.
                                               \label{Dil}
\eeq
These mappings have the points \mkb{$\{1,-1\} \in S^{1}$} as fixpoints.
The upper and lower semi-circle, \mkb{$S^{1}_{+}$}
respectively \mkb{$S^{1}_{-}$}, are mapped by dilations onto themself.
A dilation-group attached to an arbitrary, proper interval
\mkb{$\intv \subset S^{1}$} (mapping this interval onto itself) is constructed
as follows:
\begin{equation}
PSU(1,1) \ni
Dil_{\intv}(t):=g^{-1}_{\intv}Dil(t)g_{\intv},\;\;\;
   g_{\intv}\in PSU(1,1),\;\; g_{\intv}\intv = S_{+}^{1}.
                                                            \label{dil1}
\end{equation}
The interval \mkb{$ \intv $} is mapped bijectively to the upper semi-circle,
dilated and mapped back as one can see in the following diagram:\\
%
%
%
%
\[
  \putsqquare<1`1`-1`1;700`400>(-400,50)[\intv`\intv`S_{+}^{1}`S_{+}^{1};
    Dil_{\intv}`g_{\intv}`g^{-1}_{\intv}`Dil]
\]
%
%
%
%
The representation of the one-parameter group of dilations (eq. \ref{dil1}) gives
(a geo\-metric realization of) the unique modular group
\mkb{$\{ \Delta^{it}_{\intv};\: t\in \R\}$} \cite{Haag,BR} of the (vacuum) tuple
\mkb{$( \W(\intv),\omega_{0})$}.
For the case of the upper and lower semi-circle which in the non-compact
picture become positive and negative lightrays this follows from the work of
Bisognano and Wichmann \cite{BW}.
It is a peculiarity of \mkb{$CFT_{2}$} that arbitrary intervals can be mapped
onto the upper (or lower) semi-circle which therefore allows to identify the
modular group of algebras localized in these regions with the above
constructed dilations.\\
The vacuum expectation values of Weyl-operators obey the
$KMS$-condition \cite{HHW,Haag,BR} with respect to (the representation of) the
one-parameter group of dilations \cite{Y}:
\beqn
\omega_{0}(W(f)\:Ad[U_{Dil_{\intv}(t)}](W(g)))
\stackrel{\mbox{\tiny{KMS}}}{=}
     \omega_{0}(Ad[U_{Dil_{\intv}(t+i)}](W(g))\:W(f))
                                             \label{kmsvak}
\eeqn
\mkb{$W(f),\:W(g) \in \W(\intv)$}.
In the case of the vacuum state this is a necessary and sufficient condition
to identify uniquely the one-parameter group of dilations as the modular group
\cite{KR} mentioned above.\\
By a simple reparametrization of the unit-circle \mkb{$S^{1}\subset \C$} in
terms of the conformal mapping \mkb{$S^{1}\ni z \longmapsto z^{n},\; 1<n \in \N$},
Schroer and Wiesbrock construct a geometrical state \mkb{$\omega_{2}$} for the
case \mkb{$n=2$}. We shall henceforth refer to it as F-S state.
This state is shown to be invariant under
transformations of the form (eq. \ref{nnt}) for \mkb{$n=2$}:
\beq
  g_{2}(z):= \left( \frac{\alpha z^{2} + \beta}
             {\bar{\beta}z^{2}+\bar{\alpha}} \right)^{\frac{1}{2}}\!\!,
          \;\;\;\left(\matrix{
                            \alpha & \beta \cr
                       \bar{\beta} & \bar{\alpha} \cr
                            }\right) \in  PSU(1,1).
\eeq
For intervals:
\beq
\intv^{\mkb{{\tiny{$\frac{1}{2}$}}}}:=
  \intv_{1}\cup
\intv_{2},\;\;\;\:
    \intv_{i} \stackrel{z^{2}}{\longrightarrow}\intv,\;i=1,2,
                                                              \label{INT}
\eeq
the modified dilations act the following way:
\begin{eqnarray}
 Dil_{2,\intv
                           }(t)\:(\bullet) &:= &\;\;\;\;\:
   \left(Dil_{ \intv }(t)\:(\bullet)^{2}\right)^{\frac{1}{2}}
                                                          \label{dil2}
                                                                \nonumber\\
&\: = & \left(
        g^{-1}_{\intv}\:Dil(t)\:g_{\intv}\:(\bullet)^{2}
        \right)^{\frac{1}{2}}
\end{eqnarray}
(see the following diagram)\\
%
%
\[
  \putqtriangle<0`1`0;300>(-550,400)[\intv^{\frac{1}{2}}``;`(\bullet)^{2}`]
  \putsqquare<1`1`-1`1;600`400>(-250,0)[\intv`\intv`S_{+}^{1}`S_{+}^{1};
   Dil_{\intv}`g_{\intv}`g^{-1}_{\intv}`Dil]
  \putptriangle<0`0`-1;300>(350,400)[`\intv^{\frac{1}{2}}`\intv;
                                          ``(\bullet)^{\frac{1}{2}}]
  \putmorphism(-500,690)(-1,0)[``Dil_{2,\intv
                                                      }]{1100}1a
\]
%
%
\[\]\vspace{-0.24cm}
The F-S state \mkb{$\omega_{2}$} may be defined in the following manner:
\beq
 \omega_{2}(W(f)\:W(g)):=
 \omega_{0}(W(f_{\mkb{{\tiny{$\frac{1}{2}$}}}})\:
                     W(g_{\mkb{{\tiny{$\frac{1}{2}$}}}}))
 ,\;\;\;W(f),\:W(g) \in \W(\tilde{\intv})
                                                        \label{defs}
\eeq
with
\mkb{$f_{\mkb{{\tiny{$\frac{1}{2}$}}}}(\bullet):=
                             f((\bullet)^{\mkb{{\tiny{$\frac{1}{2}$}}}}),\;
supp(f_{\mkb{{\tiny{$\frac{1}{2}$}}}})  \subset \intv,\; \tilde\intv
\stackrel{z^{2}}{\longrightarrow} \intv$}. It amounts to the
following pointwise prescription for the current two-point-function:
\beqn
   \omega_{2}(J(z)J(w)):=
         2z\:2w\:
   \omega_{0}(J(z^{2})J(w^{2})).
\eeqn
At this point it is essential that the interval $\tilde{\intv}$, i.e. localization
region of the algebra $\W(\tilde\intv)$ in equation \ref{defs}, does not contain
opposite points $z, w \in S^{1}, \;arg(z)- arg(w)=0,\; mod(\frac{2\pi}{2})$.
Otherwise the F-S state becomes non-faithful as one can see easily by the following
example of an operator $W \in \W(\tilde{\intv})$:
\beq
W:=1-W(g)W(f)
\eeq
$$
f|_{supp(f)}=-g|_{supp(g)}, \;\;
supp(f)=-supp(g) \subset \tilde \intv.
$$
Localizing the algebra $\W$ in only one of the two intervals
$\intv_{i} \subset \intv^{\mkb{{\tiny{$\frac{1}{2}$}}}},\; i=1,2$,
one is able to identify the modular group of the standard tuple
\mkb{$(\W(\intv),\omega_{2})$} ($\intv$ stands for the chosen interval, i.e. $\intv_1$
respectively $\intv_2$)
with the transformations in (eq. \ref{dil2})
by showing the $KMS$-property for the state $\omega_{2}$\footnote{
Schroer and Wiesbrock did essentially the calculations in \cite{SW} whereas
in \cite{SF} the correction, i.e. the limitation to only one of the two
intervals as localization region was mentioned.}:
\begin{eqnarray}
  \omega_{2}( W(f)\:Ad[U_{Dil_{2,\intv
                                   }(t)}](W(g)))
 \stackrel{\mbox{\tiny{KMS}}}{=}
  \omega_{2}( Ad[U_{Dil_{2,\intv
                    }(t+i)}](W(g))\:W(f)).
                                                         \label{KMS}
\end{eqnarray}
Using equation \ref{defs} this may be reduced to the vacuum case:
\begin{eqnarray}
  \omega_{0}(W(f_{\mkb{{\tiny{$\frac{1}{2}$}}}})
                      \:Ad[U_{Dil_{\intv}(t)}](W(g_{\mkb{{\tiny{$\frac{1}{2}$}}}})))
  \stackrel{\mbox{\tiny{KMS}}}{=}
  \omega_{0}(Ad[U_{Dil_{\intv}(t+i)}](W(g_{\mkb{{\tiny{$\frac{1}{2}$}}}}))
                      \:W(f_{\mkb{{\tiny{$\frac{1}{2}$}}}})).
\end{eqnarray}
The faithful F-S state $\omega_{2}$ on the algebra $\W(\intv)$ has a unique
vector implementation $|\omega_{2}\rangle$ in the natural cone
$\mathcal{P}_{|\omega_{0}\rangle}$ of the standard pair $(\W(\intv),|\omega_{0}\rangle)$
\cite{SF, Haag}:
\beq
    \omega_{2}(W)=\langle\omega_{2}|W|\omega_{2}\rangle,\; W\in\W(\intv)
\eeq
We want to point out in the following section that
the above results of Schroer and Wiesbrock concerning the invariance of the
modified states with respect to the modified \mkb{$PSU(1,1)$} group and the
$KMS$-property of these states with respect to the modified dilations are
general properties for any \mkb{$CFT_{2}$} since the mentioned properties can
-as we will show- be drawn through the substitutions
\mkb{$z \longmapsto z^{n},\; \forall n \in \N$}.\\[0.3cm]
%
%
%

{\Large{\bf{III}}}
In the following we keep the pointwise prescription of fields.
We generalize to arbitrary $n \in \N$,
defining states $|\omega_{n}\rangle$ for local chiral primary fields
$\phi(z)$ on $S^{1}$ by the identity:
\beqn
  \langle \omega_{n}|
                    \prod_{k=1}^{l}
                    \phi_{k}(z_{k})
  |\omega_{n}\rangle :=
  \prod_{s=1}^{l}(nz_{s}^{n-1})^{ \Delta_{s} }
  \langle \omega_{0}|
                 \prod_{k=1}^{l}
                 \phi_{k}(z^{n}_{k})
  |\omega_{0} \rangle.
                                               \label{omen}
\eeqn
Using the M\"obius-invariance of the vacuum $|\omega_{0} \rangle$ one can show
the invariance of $|\omega_{n}\rangle$ under transformations of the form (eq.
\ref{nnt}) as follows:
\[
\phi_{i}(z_{i}) \longrightarrow
(\partial_{z} \{g_{n}(z)\}|_{z=z_{i}})^{\Delta_{i}} \phi_{i}(g_{n}(z_{i}))
\]
\[
 \langle \omega_{n}| \prod_{k=1}^{l}
                     \phi_{k}(z_{k})
                                    |\omega_{n}\rangle
 \longrightarrow
 \prod_{i=1}^{l}(\partial_{z} \{g_{n}(z)\}|_{z=z_{i}})^{\Delta_{i}}
 \:\langle \omega_{n}| \prod_{j=1}^{l}\phi_{j}(g_{n}(z_{j}))|\omega_{n}\rangle
\]
\[
\phantom{a}\;\;\;\;\;\;\;\;\;\;\;\; \stackrel{(eq. \ref{omen})}{=}
\prod_{i=1}^{l}(\partial_{z} \{g_{n}(z)\}|_{z=z_{i}})^{\Delta_{i}}
\prod_{j=1}^{l} (n\{g_{n}(z_{j})\}^{n-1})^{\Delta_{j}}
  \:\langle \omega_{0}|
         \prod_{k=1}^{l}\phi_{k}( \{g_{n}(z_{k})\}^{n} )
    |\omega_{0} \rangle
\]
\[
\phantom{a}\;\;\;\;\;\;\;\;\;\;\;\;\;\stackrel{(eq. \ref{nnt})}{=}
 \prod_{i=1}^{l}(\partial_{z} \{g_{n}(z)\}|_{z=z_{i}})^{\Delta_{i}}
 \prod_{j=1}^{l} (n\{g_{n}(z_{j})\}^{n-1})^{\Delta_{j}}
 \:\langle \omega_{0}|
         \prod_{k=1}^{l}\phi_{k}(g(z^{n}_{k}))|\omega_{0} \rangle,\;
\]
\[
  \;\;\;\;\;\;\;\;\;\;\;\;\;\;\;\;\;\;\;\;
  \;\;\;\;\;\;\;\;\;\;\;\;\;\;\;\;
  \;\;\;\;\;\;\;\;\;\;\;\;\;\;\;\;\;\;\;\;
  \;\;\;\;\;\;\;\;\;\;\;\;\;\;\;\;
  \;\;\;\;\;\;\;\;\;\;\;\;\;\;\;\;\;\;\;\;g \in PSU(1,1)
\]
\begin{eqnarray}
&\;=& \!\!\!\!\!
 \prod_{i=1}^{l}(\partial_{z} \{g_{n}(z)\}|_{z=z_{i}})^{\Delta_{i}}
 \prod_{j=1}^{l} (n\{g_{n}(z_{j})\}^{n-1})^{\Delta_{j}}
                                                \times
                                              \nonumber \\
& & \;\;\;\;\;\;\;\;\;\;\;\;\;\;\:\;\;\;\;\;\:\;
 \prod_{r=1}^{l}\frac{1}{ (\partial_{z}\{g(z)\}|_{z=z^{n}_{r}})^{\Delta_{r}}}
   \:\langle \omega_{0}|
                        \prod_{k=1}^{l}\phi_{k}(z_{k}^{n})
                       |\omega_{0} \rangle
                                                    \label{ninv}\\
&\;\stackrel{(eq. \ref{omen})}{=}& \!\!\!
  \langle \omega_{n}|
    \prod_{i=1}^{l}\phi_{i}(z_{i})
  |\omega_{n}\rangle.
                                               \nonumber
\end{eqnarray}
For the one-parameter group of modified dilations:
\beqn
Dil_{n}(t)z&:=&(Dil(t)z^{n})^{\frac{1}{n}}
                                     \nonumber\\
           &\:=&\left( \frac{ch(\pi t)z^{n}+sh(\pi t)}
                            {sh(\pi t)z^{n}+ch(\pi t)}\right)^{\frac{1}{n}}
                                              \label{diln}
\eeqn
one would like the following $KMS$-relation to be satisfied:
\beqn
 (\!\!\!\!\!&\partial_{v}&\!\!\!\!\{Dil_{n}(t)v\}|_{v=z_{l}})^{\Delta_{l}}
 \langle\omega_{n}|
                     \prod_{k=1}^{l-1}
                      \phi_{k}(z_{k})\:
                             \phi_{l}(Dil_{n}(t)z_{l})
                                     |\omega_{n}\rangle
                                                        \nonumber \\
& &\;\;\stackrel{(eq. \ref{omen})}{=}
 \prod_{i=1}^{l-1} (nz_{i}^{n-1})^{\Delta_{i}}\;
 \left[ (\partial_{v}\{Dil_{n}(t)v\}|_{v=z_{l}})^{\Delta_{l}}
 (n\{Dil_{n}(t) z_{l}\}^{n-1})^{\Delta_{l}}
 \right]
                                             \times
                                             \nonumber \\
& &\;\;\;\;\;\;\;\;\;\;\;\;\;\;\;\;
   \;\;\;\;\;\;\;\;\;\;\;\;\;\;\;\;
   \;\;\;\;\;\;\;\;\;\;\;\;\;\;\;\;\;\;\;\;\;\;\;\;\;\;\;\;
 \langle\omega_{0}|
                     \prod_{k=1}^{l-1}
                     \phi_{k}(z^{n}_{k})\:
                         \phi_{l}(Dil(t)z^{n}_{l})
                   |\omega_{0}\rangle
                                                        \nonumber \\
& &\;\;\:=\;\;\;
 \prod_{i=1}^{l} (nz_{i}^{n-1})^{\Delta_{i}}\;
 (\partial_{v}\{Dil(t)v\}|_{v=z^{n}_{l}})^{\Delta_{l}}
              \langle\omega_{0}|
                               \prod_{k=1}^{l-1}
                               \phi_{k}(z^{n}_{k})\:
                                   \phi_{l}(Dil(t)z^{n}_{l})
                               |\omega_{0}\rangle
                                                        \nonumber \\
& &\!\! {! \atop \stackrel{ \mkb{ {\tiny{$(KMS)$}} } }{=}}
 \prod_{i=1}^{l} (nz_{i}^{n-1})^{\Delta_{i}}\;
 (\partial_{v}\{Dil(t+i)v\}|_{v=z^{n}_{l}})^{\Delta_{l}}
                                    \times
                                    \nonumber \\
& &\;\;\;\;\;\;\;\;\;\;\;\;\;\;\;\;
   \;\;\;\;\;\;\;\;\;\;\;\;\;\;\;\;
   \;\;\;\;\;\;\;\;\;\;\;\;\;\;\;\;\;\;\;\;\;
              \langle\omega_{0}|
                                \phi_{l}(Dil(t+i)z^{n}_{l})\:
                                    \prod_{k=1}^{l-1}
                                    \phi_{k}(z^{n}_{k})
                               |\omega_{0}\rangle
                                                         \nonumber\\
& &\;\stackrel{(eq. \ref{omen})}{=}
(\partial_{v}\{Dil_{n}(t+i)v\}|_{v=z_{l}})^{\Delta_{l}}
\langle\omega_{n}|
                  \phi_{l}(Dil_{n}(t+i)z_{l})\:
                      \prod_{k=1}^{l-1}
                      \phi_{k}(z_{k})
                  |\omega_{n}\rangle.
                                                         \label{mkks}
\eeqn
We have managed here to prove the \mkb{$KMS$}-condition for the chiral part of
the correlation function in the modified \mkb{$(|\omega_{0}\rangle \mapsto
|\omega_{n}\rangle)$} theory provided that the \mkb{$KMS$}-condition holds in
the unmodified theory. But the later is not precisely true.
One picks in general a monodromy shift in the analytic continuation
\mkb{$t \mapsto t+is,\; s \in [0,1]$}
\footnote{
In statistical physics one has the interval \mkb{$[0,\beta]$}, \mkb{$\beta$}
the inverse temperature, and the states are called $\beta$-$KMS$-states
\cite{Haag,BR}.
          },
which amounts to a full circle in complex space -returning to the same point
on a different Riemann sheet. We may assume that the monodromy under
consideration is diagonalized in a suitably chosen basis of conformal blocks
and that the ensuing phase factors are compensated by the inverse phase
factors of the anti-chiral block function (we assume the
\mkb{$\phi_{(i,j)}(z,\bar{z})$} to be scalar operators). The general
situation is, what concerns the cancellation of phasefactors, faithfully
represented by the simplified situation in the case of two-point-functions.
The (vacuum-) $2$-point-functions are already defined by conformal invariance
(and locality: \mkb{$\bar{\Delta}=\Delta$}) to be of the form \cite{DiF}:
\beq
   \langle\omega_{0}|\phi_{1}(z,\bar{z})
                     \phi_{2}(w,\bar{w})
                     |\omega_{0}\rangle =
                                        \frac{C_{12}}{|z-w|^{4\Delta}}.
                                                        \label{u1}
\eeq
This $2$-point-function fulfills the $KMS$-condition with respect to the
one-parameter group of dilations \mkb{$Dil(t) \in PSU(1,1)$} (eq. \ref{Dil}) as
one can show by direct calculation. In the case of arbitrary
$n$-point-functions of local conformal fields \mkb{$\phi_{(i,j)}(z,\bar{z})$}
the above argument which holds for a rational conformal field theory
\cite{DiF} reduces the analytic structure essentially to the
$U(1)$-vertex-form, i.e. equation \ref{u1}. Since the $KMS$-property holds
for (eq. \ref{u1}) it follows for the $n$-point-functions as well. Here it is
important that one needs both, the chiral and anti-chiral part.\\
One has the following $KMS$-condition for general local conformal fields:
\beqn
(\!\!\!\!\!&\partial&\!\!\!\!\!_{v}\{Dil(t)v\}|_{v=z_{l}})^{\Delta_{i_{l}}}
(\partial_{\bar{u}}\{Dil(-t)\bar{u}\}
                  |_{\bar{u}=\bar{z}_{l}})^{\Delta_{j_{l}}}
                                  \times
                                                        \nonumber \\
& & \;\;\;\; \;\;\;\;
\langle\omega_{0}|
                     \prod_{k=1}^{l-1}
                     \phi_{(i_{k},j_{k})}(z_{k},\bar{z}_{k})\:
                     \phi_{(i_{l},j_{l})}(Dil(t)z_{l},Dil(-t)\bar{z}_{l})
                                 |\omega_{0}\rangle
                                                        \nonumber \\
& & \stackrel{\mkb{{\tiny{$KMS$}}}}{=}
(\partial_{v}\{Dil(t+i)v\}|_{v=z_{l}})^{\Delta_{i_{l}}}
(\partial_{\bar{u}}\{Dil(-t-i)\bar{u}\}
                  |_{\bar{u}=\bar{z}_{l}})^{\Delta_{j_{l}}}
                                  \times
                                                        \nonumber \\
& & \;\;\;\; \;\;\;\; \;\;\;\;
\langle\omega_{0}|
                   \phi_{(i_{l},j_{l})}(Dil(t+i)z_{l},Dil(-t-i)\bar{z}_{l})\:
                   \prod_{k=1}^{l-1}
                   \phi_{(i_{k},j_{k})}(z_{k},\bar{z}_{k})\:
                           |\omega_{0}\rangle.
                                                        \nonumber \\
\eeqn
A sufficient condition for locality is \mkb{$\Delta_{i_{p}}=\Delta_{j_{p}},\;
p=1,\dots,l$}.

Going now to the Schroer-Wiesbrock ansatz, one has the
following picture on the chiral and anti-chiral sector,
respectively, and therefore the modified states \mkb{$|\omega_{n,n}\rangle$}:
\beqn
  \langle \omega_{n,n}|
                             \prod_{k=1}^{l}
                    \!\!\!\!&\phi&\!\!\!\!_{(i_{k},j_{k})}(z_{k},\bar{z}_{k})
  |\omega_{n,n}\rangle :=
                                                        \nonumber \\
& &
  \prod_{s=1}^{l}(nz_{s}^{n-1})^{ \Delta_{i_{s}} }
  \prod_{q=1}^{l}(n\bar{z}_{q}^{n-1})^{ \Delta_{j_{q}} }
  \langle \omega_{0}|
                 \prod_{k=1}^{l}
                 \phi_{(i_{k},j_{k})}(z^{n}_{k},\bar{z}^{n}_{k})
  |\omega_{0} \rangle.
                                                        \label{mnn}
\eeqn
Again, the $KMS$-condition with respect to
\mkb{$Dil_{n}(t)z:=(Dil(t)z^{n})^{\frac{1}{n}}$} (eq. \ref{diln}) on both sectors
''goes through \mkb{covariantly}'' as in equation (eq. \ref{mkks}).\\
An illustration of what had been said above in a non-trivial setting (the
$U(1)$-current-algebra is a quasifree theory \cite{BMT}) is provided by the
theory of non-abelian currents \cite{DiF}. The central relation is the
current-current commutation relation (Kac-Moody-algebra, with the circle as
base space):
\beq
          [J^{a}(z),J^{b}(w)]=if^{abc}J^{c}(z)\delta(z-w)-
                     k\delta^{ab}\partial_{z}\delta (z-w).
                                                        \label{kac}
\eeq
This relation allows it to calculate the $m$-point correlation function
recursively by using the \mkb{$m$-$1$}- and \mkb{$m$-$2$}-point-function:
\beqn
  \langle \omega_{0}|
                     \prod_{i=1}^{m} J^{a_{i}}(z_{i})
  |\omega_{0}\rangle
   &\stackrel{(eq. \ref{kac})}{=}&
      \sum_{j=2}^{m}
           \frac{k \delta^{a_{1}a_{j}}}{ (z_{1} - z_{j})^{2} }
     \: \langle \omega_{0}|
                 \prod_{k=2 \atop k\ne j}^{m} J^{a_{k}}(z_{k})
         |\omega_{0}\rangle
            +
                                                       \nonumber \\
   & &
      \sum_{j=2}^{m}
          \frac{i f^{a_{1}a_{j}d} }{(z_{1} - z_{j})}
       \:
          \langle \omega_{0}|
                 \prod_{k=2}^{j-1} J^{a_{k}}(z_{k})
                           J^{d}(z_{j})
               \!\!  \prod_{l=j+1}^{m}\!\!  J^{a_{l}}(z_{l})
          |\omega_{0}\rangle.
                                                       \nonumber \\
                                                       \label{npkt}
\eeqn
One can therefore verify the $KMS$-property inductively since the $2$-point
function has the $KMS$-property. For the \mkb{$|\omega_{n}\rangle$} state one
gets the following recurrence relation:
\beqn
  \langle \omega_{n}|
                     \prod_{i=1}^{m} J^{a_{i}}(z_{i})
  |\omega_{n}\rangle
    &\stackrel{(eq. \ref{omen})}{=}&
         \sum_{j=2}^{m}
           \frac{(nz^{n-1}_{1})\:(nz^{n-1}_{j})\: k\delta^{a_{1}a_{j}}}
                {(z^{n}_{1} - z^{n}_{j})^{2}}
          \: \langle \omega_{n}|
                 \prod_{k=2 \atop k\ne j}^{m} J^{a_{k}}(z_{k})
           |\omega_{n}\rangle
          +
                                                      \nonumber \\
    & &\!\!\!\!\!\!\!\! \!\!\!\!\!\!\!\!
         \sum_{j=2}^{m}
           \frac{(nz^{n-1}_{1})\:i f^{a_{1}a_{j}d} }
               {(z^{n}_{1} - z^{n}_{j})}
       \:
          \langle \omega_{n}|
                 \prod_{k=2}^{j-1} J^{a_{k}}(z_{k})
                           J^{d}(z_{j})
               \!\!  \prod_{l=j+1}^{m}\!\!  J^{a_{l}}(z_{l})
          |\omega_{n}\rangle.
                                                      \nonumber
\eeqn
The $KMS$-condition with respect to
\mkb{$Dil_{n}(t)z:=(Dil(t)z^{n})^{\frac{1}{n}}$} (eq. \ref{diln}) can be shown
inductively as well, since the substitution \mkb{$z \longmapsto z^{n}$} goes
through covariantly and therefore the $KMS$-property for the vacuum carries
over to the new state.\\
For the identification with the modular group mentioned above one needs
von Neumann algebras. It is a subtle problem to proceed from smeared unbounded
field operators to local algebras of bounded operators. By using bounded
functions of unbounded local operators as in the case of the
$U(1)$-current-algebra one has to ensure locality which might be not conserved
by this mapping. Bisognano and Wichmann gave certain sufficient conditions to
identify the von Neumann algebras to which the (Wightman-)
field-algebras are \mkb{affiliated} \cite{BW,BR}. Generally it is a
non-trivial task to verify these conditions. In the case of non-abelian
currents, this is possible and one can therefore speak of von Neumann
algebras, respectively the associated \mkb{modular} group.\\[0.3cm]
%
%

{\Large{\bf{IV}}}
We have seen that the localization region of the algebras has to exclude opposite
points to guarantee the faithfulness of the F-S state $\omega_{n}$.
In the following we will stick to the case $n=2$ for simplicity. Since both
intervals $\intv_{1,2}$ (eq. \ref{INT}) are allowed for localizing the algebra
it seems natural to have a closer look to the two-interval, i.e. multilocalized,
algebra $\W(\intv_{1}) \vee \W(\intv_{2})=\W(\intv_{1} \cup \intv_{2})
=:\W(\intv^{\mkb{{\tiny{$\frac{1}{2}$}}}})$.
The demand for faithfulness prohibits the direct use of the F-S state $\omega_{2}$.
There is however a way to enforce faithfulness by applying the split-property
\cite{Haag,Sch}\footnote{Another way might be the use of a (minimal)
projector $E$ implying the faithfulness of $\omega_2$ on the reduced algebra
$E\W E$.
}.\\
For two spacelike regions, i.e. disjoint intervals, the split-property implies
the following:
\beq
 \W(\intv_{1}) \vee \W(\intv_{2}) \simeq \W(\intv_{1}) \otimes \W(\intv_{2}).
 \label{nw}
\eeq
This reflects in a certain sense the statistical independence  of the algebras.
The state $\omega_{2}$, faithful and normal over either interval $\intv_{1,2}$ can
be extended to a product state, faithful and normal over
$\W(\intv^{\mkb{{\tiny{$\frac{1}{2}$}}}})$, i.e.:
\beq
  \omega_{2}^{p}(WV)= \omega_{2}(W)  \omega_{2}(V),\;\;
  W \in \W(\intv_{1}),\; V \in \W(\intv_{2})
\eeq
The modular group also splits and one can show the following theorem:\\

{\bf{Theorem}}: The modular group $\sigma^{t}_{\omega_{2}^{p}}$ of the faithful
product state $\omega_{2}^{p}$ on the algebra
$\W(\intv^{\mkb{{\tiny{$\frac{1}{2}$}}}})$ is given by the geometric action of
$Dil_{2}(t)$. Moreover the unitary implementer $U_{Dil_{2}(t)}$ whose
infinitesimal generator is a linear combination of $L_{\pm 2}$
\beq
\sigma^{t}_{\omega_{2}^{p}}=Ad[U_{Dil_{2}(t)}]
\label{SigmaDil2}
\eeq
is the $\Delta^{it}_{\omega_{2}^{p}}$ modular object of the state vector
$|\eta_{2}\rangle$ which represents the faithful product state $\omega_{2}^{p}$
in the positive cone of
$(\W(\intv^{\mkb{{\tiny{$\frac{1}{2}$}}}}),|\omega_{0}\rangle)$.\\[0.2cm]
Since the arguments are entirely similar to those which demonstrated that
$U_{Dil_{2}(t)}$ was the $\Delta^{it}$ modular group of the faithful state $\omega_2$
on either $\W(\intv_{1})$ or $\W(\intv_{2})$, we omit the details. The crucial point
is the invariance:
\beqn
\omega_{2}^{p}(Ad[U_{Dil_{2}(t)}](WV)) &=& \omega_{2}^{p}(Ad[U_{Dil_{2}(t)}](W)
\:Ad[U_{Dil_{2}(t)}](V))\nonumber \\
                    &=& \omega_{2}(Ad[U_{Dil_{2}(t)}](W))
                      \:\omega_{2}(Ad[U_{Dil_{2}(t)}](V))\nonumber\\
                    &=& \omega_{2}(W) \:\omega_{2}(V)\nonumber\\
                    &=& \omega_{2}^{p}(WV)
\eeqn
which uses the previously established invariance of $\omega_{2}$. The aforementioned
lack of faithfulness of the state $\omega_{2}$ on the multi-interval algebra
$\W(\intv^{\mkb{{\tiny{$\frac{1}{2}$}}}})$ is related to the geometric  nature of the
double interval modular group $\sigma^{t}_{\omega_{2}^{p}}$.\\
The remarkable fact that the pairs $(\W(\intv_{i}), \omega_{2}),\; i=1,2,$
$(\W(\intv^{\mkb{{\tiny{$\frac{1}{2}$}}}}),\omega_{2}\times \omega_{2})$ share the
same modular group action and (in the appropriate positive cones) have the same
implementing modular unitaries is also related to this. The factorization of the
$\omega_{0}$ vacuum would not lead to such a situation.\\
This result raises the question whether both modular objects of the double
interval situation $(\W(\intv^{\mkb{{\tiny{$\frac{1}{2}$}}}}),|\eta_{2}\rangle)$
can be geometric. Formally the candidate for a geometric $J$ is the "rotated" TCP
transformation $z \rightarrow -\bar{z}$ which maps for instance the first
quarter-circle $S^{1}_{1}:=[0,\frac{\pi}{2}]$ to the second quarter-circle
$S^{1}_{2}:=[\frac{\pi}{2},\pi]$ and the third quarter-circle $S^{1}_{3}$ to the
fourth $S^{1}_{4}$ \footnote{The quarter-circle algebras $\W(S^{1}_{1,3})$ related to
the algebra localized on the upper semi-circle $\W(S^{1}_{+})$ provide
a natural example.}
(this transformation has the same geometric effect as the
analytically continued $\Delta^{\frac{1}{2}}_{\omega_{2}^{p}}$).
However, the replacement of Haag duality by an inclusion:
\beq
\W((\intv_{1} \cup \intv_{2})') \subset \W(\intv_{1} \cup \intv_{2})'
\eeq
and its explanation in term of superselection sectors shows that the true
modular involution has in addition to the geometric part an algebraic
modification. A calculation of $J$ for abelian current models seems to be feasible.
Via the inverse algebraic lightfront holography \cite{SF} these geometric chiral
transformations correspond to fuzzy symmetries in the original net.\\
%
%
%

{\Large{\bf{V}}}
It was shown that $Dil_{n}$ (having $2n$ fixpoints instead of two as $Dil$)
is the modular group of the standard tuple $(\W(\intv),\omega_n)$.
The interval $\intv$ is forbidden to contain points
$z, w \in S^{1}, \;arg(z)- arg(w)=0,\; mod(\frac{2\pi}{n})$.\\
With regard to the modular origin of the Witt-Virasoro basis it is
sufficient to extend the construction up to the $n=2$ case due to the
relation (\ref{Vir}).\\
In the case of multi-interval, i.e. multilocalized Weyl-algebras (eq. \ref{nw}),
the transformations \mkb{(eq. \ref{SigmaDil2})}
can be identified with the modular group of the tuple
\mkb{$(\W(\intv^{\mkb{{\tiny{$\frac{1}{2}$}}}}),|\eta_{2}\rangle)$}.
Going to higher $n$ by using the split-property is not totally straightforward.\\
Following the program of algebraic quantum field theory this result underlines
the special r\^ole played by modular theory in general local quantum field
theory.
Schroer and Wiesbrock and Schroer and Fassarella \cite{SF} suggest using
modular theory as a tool to explore ''fuzzy'' symmetries i.e. symmetries which
do not originate from the classical Noether setting.
They also propose to investigate the relation of their findings with the notion of
{\emph{halfsided-modular-inclusion(intersections)}} \cite{W1,B2,W2}.
See \cite{DB,KW} for a recent account and further references to the results in
\cite{W1,B2,W2}.\\
To investigate the latter was out of scope of the present work.
Here, we showed that a new ansatz of Schroer and Wiesbrock (corrected in \cite{SF})
can be dealt with from a more general point of view, since the $KMS$-property
of the vacuum state (a general property for local fields in the
vacuum-setting) carries over covariantly to the set of new F-S states.
The argument for the general validity of the $KMS$-property given above is
true for rational conformal field theories.\\
Multilocalized algebras can be defined by using the split-property which implies
the faithful states to be a product of the new constructed F-S states.
The modular group naturally splits into a product of two copies of the
modified dilations whereas the modular conjugation needs some further
investigation. \\[0.6cm]
%

{\emph{Note to be added}}: We are indebted to Prof. Schroer for pointing out to
us the incorrectness of the implicitly assumed faithfulness of the
new states with respect to multilocalized algebras in \cite{SW}.
Section 4 is based on notes by Prof. Schroer.

{\bf Acknowledgements:}
The author is grateful to Prof. Dr. R. Flume for suggesting this work.
Also,we would like to thank Kayhan \"Ulker and Nicolai von Rummell for
help and wise comments when preparing this work. I thank also H. Gutmann
for a careful reading of the manuscript.\\

\end{document}